\documentclass[%
pra,
reprint,
superscriptaddress,
nofootinbib,
amsmath,
amssymb,
aps,
]{revtex4-1}

\usepackage{graphicx}
\usepackage{bm}

\usepackage{tikz}
\usetikzlibrary{calc}
\usetikzlibrary{hobby}
\usetikzlibrary{external}

\tikzset{
 pics/blob/.style={
   code={
   \draw[use Hobby shortcut, fill, closed] (0,0) +($(0.5:rnd)$)
       \foreach \a in {60,120,...,350} {  .. +($(\a:rnd)$) };
   }
}}

\usepackage[xcolor,final]{changes}
\definechangesauthor[color= blue]{gs}
\definechangesauthor[color= red]{pb}

\begin{document}

\preprint{APS/123-QED}

\title{Full characterization of the transmission properties of a multi-plane light converter}

\author{Pauline Boucher}
\affiliation{%
Laboratoire Kastler Brossel, Sorbonne Universit\'e, ENS-Universit\'e PSL,
Coll\`ege de France, CNRS; 4 place Jussieu, F-75252 Paris, France
}%
\affiliation{%
Cailabs, 38 boulevard Albert 1er, 35200 Rennes, France
}%
\author{Arthur Goetschy}
\affiliation{%
ESPCI Paris, PSL University, CNRS, Institut Langevin, 1 rue Jussieu, F-75005 Paris, France.\
}%
\author{Giacomo Sorelli}
\affiliation{%
Laboratoire Kastler Brossel, Sorbonne Universit\'e, ENS-Universit\'e PSL,
Coll\`ege de France, CNRS; 4 place Jussieu, F-75252 Paris, France
}%
\affiliation{
DEMR, ONERA, Université Paris Saclay, F-91123, Palaiseau, France
}
\affiliation{Sorbonne Universit\'e, CNRS, LIP6, 4 place Jussieu, F-75005 Paris, France}
\author{Mattia Walschaers}
\affiliation{%
Laboratoire Kastler Brossel, Sorbonne Universit\'e, ENS-Universit\'e PSL,
Coll\`ege de France, CNRS; 4 place Jussieu, F-75252 Paris, France
}%
\author{Nicolas Treps}
\affiliation{%
Laboratoire Kastler Brossel, Sorbonne Universit\'e, ENS-Universit\'e PSL,
Coll\`ege de France, CNRS; 4 place Jussieu, F-75252 Paris, France
}%

\date{\today}

\begin{abstract}
Multi-plane light conversion (MPLC) allows to perform arbitrary transformations on a finite set of spatial modes with no theoretical restriction to the quality of the transformation. 
Even though the number of shaped modes is in general small, the number of modes transmitted by an MPLC system is extremely large.
In this work, we aim to characterize the transmission properties of a multi-plane light converter inside and outside the design-modes subspace.
We report, for the first time, the construction of the full transmission matrix of such systems. By performing singular value decompositions, we individuate new ways to evaluate their efficiency in performing the design transformation. 
Moreover, we develop an analytical random matrix model that suggests that in the regime of a large number of shaped modes an MPLC system behaves like a random scattering medium with limited number of controlled channels.
\end{abstract}


\maketitle


\section{Introduction}
\label{Sec:intro}
The ability of shaping light's spatial profile is crucial for several different technologies such as imaging through opaque media \cite{Vellekoop:15} and biological tissues \cite{YU2015632}, classical \cite{schwartz2009} and quantum \cite{Sorelli_2019} communication, and quantum information processing \cite{Defienne:2016}. 
One of the first methods that has been used to manipulate light's spatial distribution was adaptive optics \cite{tyson}, which uses deformable mirrors for the real-time correction of turbulence-induced phase distortions.
Light fields' spatial profile can also be manipulated via wave-front shaping in complex scattering media \cite{Rotter_Gigan,Matthess:2019}.
In this context, the propagation medium supports a very large number of spatial modes, and couples them with one another in a complex, but static, way. 
This fact can be exploited to engineer the phase front of the incident light in order to obtain the desired output spatial distribution.

Another way to shape spatial modes of light is to control the medium they propagate through, as it happens, e.g., in complex nanostructures \cite{su2018inverse} and in photonic lanterns \cite{Birks:15,leon20130}.
In the latter, an array of single-mode fibers is gradually merged into a multimode wave guide such that the modes of the fibers are adiabatically mapped into the modes of the wave guide. 
The propagation medium can be controlled dynamically as well, using, for instance, mechanical deformations in fibers in the optical regime \cite{Resisi:2020}, or tunable metasurfaces in the microwave regime \cite{Kaina:2014aa}.

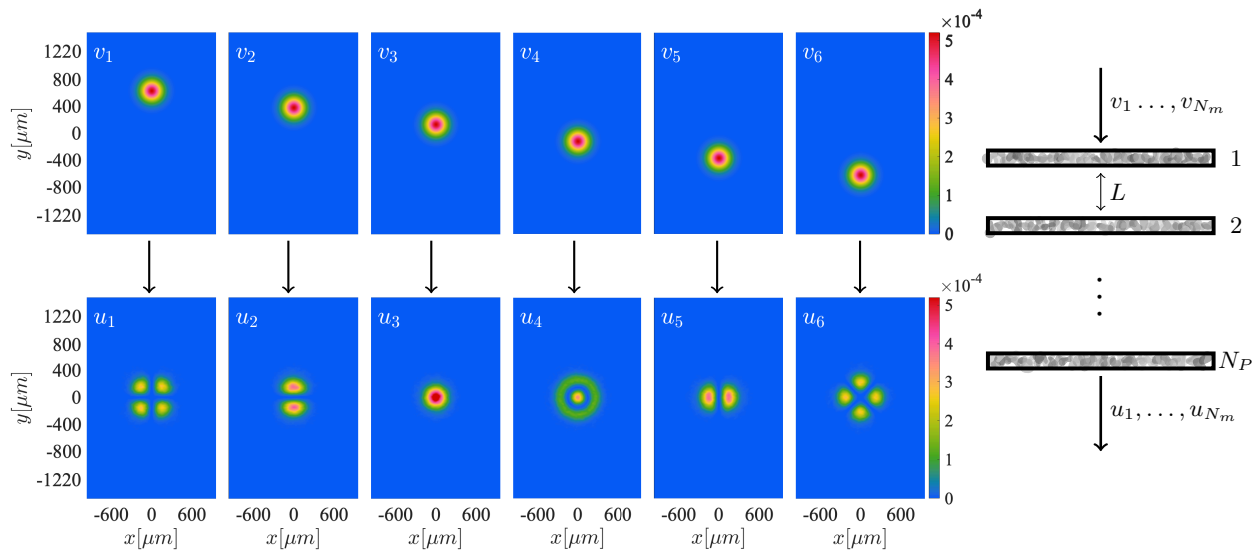
\begin{figure*}[ht]
\begin{tikzpicture}
\node at (0,0) {\includegraphics[width=0.95\textwidth]{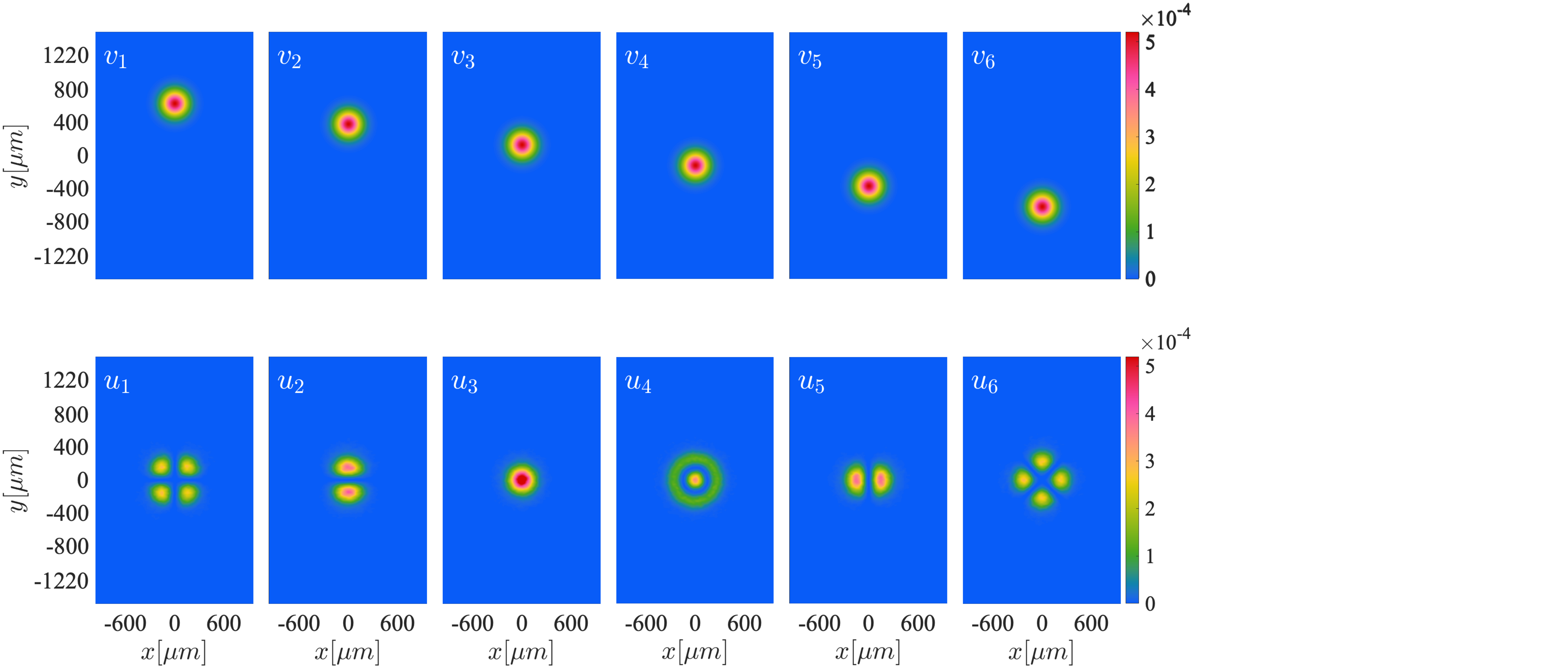}};
\draw[->, thick] (-7.45+0.8,0.5) -- (-7.45+0.8,-0.2);
\draw[->, thick] (-7.5+0.8+1.9,0.5) -- (-7.5+0.8+1.9,-0.2);
\draw[->, thick] (-7.5+0.8+2*1.9,0.5) -- (-7.5+0.8+2*1.9,-0.2);
\draw[->, thick] (-7.5+0.8+3*1.9,0.5) -- (-7.5+0.8+3*1.9,-0.2);
\draw[->, thick] (-7.5+0.8+4*1.9,0.5) -- (-7.5+0.8+4*1.9,-0.2);
\draw[->, thick] (-7.5+0.8+5*1.9,0.5) -- (-7.5+0.8+5*1.9,-0.2);
\foreach \y in {-1.1} {
\foreach \x in {4.55,4.6,...,7.45} {
    \pgfmathsetmacro{\scale}{0.09+0.05*rnd}
     \pgfmathparse{0.5*rnd+0.3}
     \definecolor{MyColor}{rgb}{\pgfmathresult,\pgfmathresult,\pgfmathresult}
    \pic at (\x,\y)  [draw= MyColor, fill=MyColor, scale=\scale, rotate=360*rnd, opacity= 0.5]{blob};
  }
};

\foreach \y in {0.7} {
\foreach \x in {4.55,4.6,...,7.45} {
    \pgfmathsetmacro{\scale}{0.09+0.05*rnd}
     \pgfmathparse{0.5*rnd+0.3}
     \definecolor{MyColor}{rgb}{\pgfmathresult,\pgfmathresult,\pgfmathresult}
    \pic at (\x,\y)  [draw= MyColor, fill=MyColor, scale=\scale, rotate=360*rnd, opacity= 0.5]{blob};
  }
};

\foreach \y in {1.6} {
\foreach \x in {4.55,4.6,...,7.45} {
    \pgfmathsetmacro{\scale}{0.09+0.05*rnd}
     \pgfmathparse{0.5*rnd+0.3}
     \definecolor{MyColor}{rgb}{\pgfmathresult,\pgfmathresult,\pgfmathresult}
    \pic at (\x,\y)  [draw= MyColor, fill=MyColor, scale=\scale, rotate=360*rnd, opacity= 0.5]{blob};
  }
};
\draw[line width = 1.,->] (6,2.8) -- node[right] {$v_1 \dots, v_{N_m}$} (6,1.8);
\draw[line width = 1.5] (4.5,1.7) rectangle (7.5,1.5);
\draw[line width = 1.5] (4.5,0.8) rectangle (7.5,0.6);
\node[rotate=90] at (6,-0.2) {\Large $\cdots$};
\draw[line width = 1.5] (4.5,-1) rectangle (7.5,-1.2);
\draw[line width = 1.,->] (6,-1.3) -- node[right] {$u_1, \dots, u_{N_m}$} (6,-2.3);
\node at (7.8,1.6) {$1$}; 
\node at (7.8,0.7) {$2$}; 
\node at (7.8,-1.1) {$N_P$}; 
\draw[<->] (6,1.4) -- node[right]{$L$} (6,0.9); 

\end{tikzpicture}
\caption{Example of an MPLC transformation that maps $N_m=6$ input modes $\{ v_i\}_{1 \leq i \leq N_m=6}\}$ (Gaussian spots) into $N_m=6$ output modes $\{ u_i\}_{1 \leq i \leq N_m=6}\}$ (fiber modes). The transformation is implemented via a series of $N_P=14$ phase plates separated by free-space propagation over a distance $L$. The reader should notice that modes $\{ u_i\}_{1 \leq i \leq N_m=6}\}$ in the bottom row are not the ideal design modes, but were obtained by (numerically) propagating the input modes through the actual device.}
\label{fig:modelisationMPLC}
\end{figure*}

Multi-plane light conversion (MPLC) is a light-shaping technique that allows to map a set of input spatial modes of light into a set of output modes by alternating free-space propagation and phase modulations (see Fig. \ref{fig:modelisationMPLC}) \cite{Morizur:10}.
MPLC systems can be designed using so-called wave-front-matching techniques \cite{morizur2019,Fontaine2019} to determine the phase modulations necessary to perform a specific mode transformation.
Such, generally complex, phase transformations are physically implemented via reflecting phase plates \cite{Morizur:10}.
A particularity of MPLC systems is that the number of input scattering channels is much larger than the number of shaped modes. 
Because of the complexity of the phase pattern on each phase plate, an MPLC system is thus expected to behave as an open chaotic cavity, producing a speckle pattern at the output for most input channels except the ones it is designed for.

So far, the study of MPLC focused on the design of specific transformations involving a certain number of modes either via optimization algorithms using a reasonably small number of phase plates \cite{Morizur:10,Labroille:14,morizur2019,Fontaine2019,Brandt:20} or via exact analytical methods using a very large number of phase transformations \cite{lopezpastor2019}.
In this work, we do not aim to present new methods for the efficient design of MPLC systems, but rather to provide a comprehensive description of the transmission properties of existing devices, and in particular of their behaviour outside the subset of  modes that they are designed to shape. 
Apart from its fundamental interest, this characterization has a practical relevance. 
In fact, construction defects and experimental imperfections (e.g. misalignment, modal crosstalk, etc.) --- which must be taken into account for an optimal use of physical devices --- lead to the injection of modes different from the design ones into the MPLC system.

By using a singular value decomposition of the transmission matrix of realistic MPLC devices in Sec. \ref{Sec:transmission}, we show how the design mode subspace and the set of modes associated to the largest singular values are related. 
A clear threshold is identified beyond which singular modes produce speckle patterns at the output. 
This analysis reveals new ways to assess MPLC systems' efficiency and suggests that, outside the design subspace, such devices behave like random scattering media.
In Sec. \ref{Sec:FRM}, we confirm this idea by deriving an analytical random matrix model which predicts the transmission eigenvalue distribution of MPLC systems. 
Finally, Sec. \ref{Sec:conclusion} concludes our work.

\section{Transmission properties of MPLC systems}
\label{Sec:transmission}

We now set out to fully characterize the transmission properties of some specific MPLC systems that have been designed (and physically constructed) by the company Cailabs, matching particular industrial requirements that we detail in Sec. \ref{Sec:definition}.
To this goal, we numerically propagate a basis of input modes through the successions of phase plates that define some particular MPLC systems. 
We then project the transmitted modes on an output-mode basis in order to reconstruct the transmission matrix of the device.
The singular values and singular vectors of the latter fully describe the transmission properties of the system.

\subsection{Definition of the MPLC systems}
\label{Sec:definition}
The operation implemented by an MPLC system is a mode-basis change, which is characterized by the number of basis elements $N_m$ and their input and output spatial profiles, which we label as $\{v_i\}_{1\leq i \leq N_m}$ and $\{u_i\}_{1\leq i \leq N_m}$ respectively (see Fig. \ref{fig:modelisationMPLC} for a specific example).
Such a transformation is performed by transmitting light trough a specific set of $N_p$ phase plates of size $m_x\times m_y$ pixels, placed at a distance $L$ from one another as sketched on the right of Fig. \ref{fig:modelisationMPLC}.

The phase profiles of the phase-plates are computed by a deterministic optimization algorithm that takes as input the details of the mode-basis change described above.
Two metrics are taken into account when designing a system.
The first metric the design algorithm tries to maximize is how close to the ideal mode basis change the transformation we implement is --- that is how well the shaped modes overlap with the design ones. 
The second figure of merit that the algorithm considers is the crosstalk between modes. 
For an ideal mode-basis change, all crosstalk coefficients are equal to zero.
However, imperfections in the transformation introduce non-zero coefficients.
In many applications for which MPLC systems are used, such as telecommunications and metrology, crosstalk between modes is an important source of errors.
Accordingly, the design algorithm tries to minimize modal crosstalk.
Another design characteristic of these systems is the set of optimization constraints taken into account at the phase plate design level. 
Indeed, all these systems are designed in an industrial setting with the goal of being physically implemented. 
This set of constraints aims at matching the physical characteristics of the numerically generated phase plates with the available manufacturing capabilities. 

The above mentioned criteria are common to all the MPLC systems analysed in this work. 
New criteria (different from the design ones) to evaluate the performance of an MPLC device will emerge from the transmission matrix analysis presented in the following sections.
However, let us stress again that our study aims at analysing these existing systems with a novel perspective to fully understand their transmission properties --- not at modifying their design.

\subsection{Construction of the MPLC transmission matrix}

We now construct the transmission matrices for a zoo of MPLC systems, mapping series of spatially separated Gaussian spots into different types of modes (free-space modes, fiber modes, etc.).
We restrict ourselves to the study of devices that have been physically constructed according to the criteria specified in Sec. \ref{Sec:definition}, and for which the validity of the prediction of our numerical propagation routine has already been verified experimentally.

The transmission matrix $t$ of an optical device maps $P$ input modes $\{\phi_i\}_{1\leq i \leq P}$ into $Q$ output modes $\{\psi_i\}_{1 \leq i \leq Q}$ according to 
\begin{equation}
\psi_i = \sum_{j=1}^P t_{ij}\phi_j.
\label{t_def}
\end{equation}
As the output mode basis $\{\psi_i\}_{1 \leq i \leq Q}$, we choose the pixel basis, for which $Q = m_x\times m_y$ is the number of pixels of the actual phase plates. 
This is a natural choice, since this is the basis used by the phase-matching algorithm to determine the phase plates of a particular MPLC system.

It is tempting to choose the pixel basis also for the input modes $\{\phi_i\}_{1\leq i \leq P}$.
However, for typical MPLC systems $Q$ is fairly large ($150\,000\leq Q\leq 400\,000$) and a matrix of size $Q \times Q$ would be numerically intractable.
We therefore chose a mode basis for which a limited number $P \ll Q$ of modes can accurately describe the input of the system.
A basis satisfying this requirement is constituted by the Hermite-Gauss (HG) modes, which are constructed as a product of $n_x$ and $n_y$ modes in the $x$ and $y$ directions, meaning $P=n_x\times n_y$. 
In particular, in our numerical simulations we considered $P= 645$ ($n_x = 15$ and $n_y=43$).

Our choice for the input-mode basis is justified by the fact that, often, the inputs of an MPLC system are spatially separated Gaussian modes. 
Because of experimental imperfections (e.g. misalignment) and construction defects (e.g. errors in positioning of the phase plates), in practice, the spatial parameters (displacement, tilt, waist size, defocus) of these modes will be altered. 
Such modified Gaussian modes can be well approximated by a linear combination of a small number of HG modes.
On the other hand, we have no a priori information on the output modes of a misaligned MPLC device, but we have experimental evidences that they resemble speckle patterns.
The high spatial resolution necessary to accurately describe such patterns is guaranteed by our choice of representing the output field with a large number of pixel modes.

Finally, to ensure that our numerical representation of the transmission matrix $t$ is accurate, we tested different types of mode bases and of mode-bases sizes without spotting any notable difference.

\subsection{Singular value decomposition}
\label{Sec:SVD}
\begin{figure}[b]
\centering
\includegraphics[width=\columnwidth]{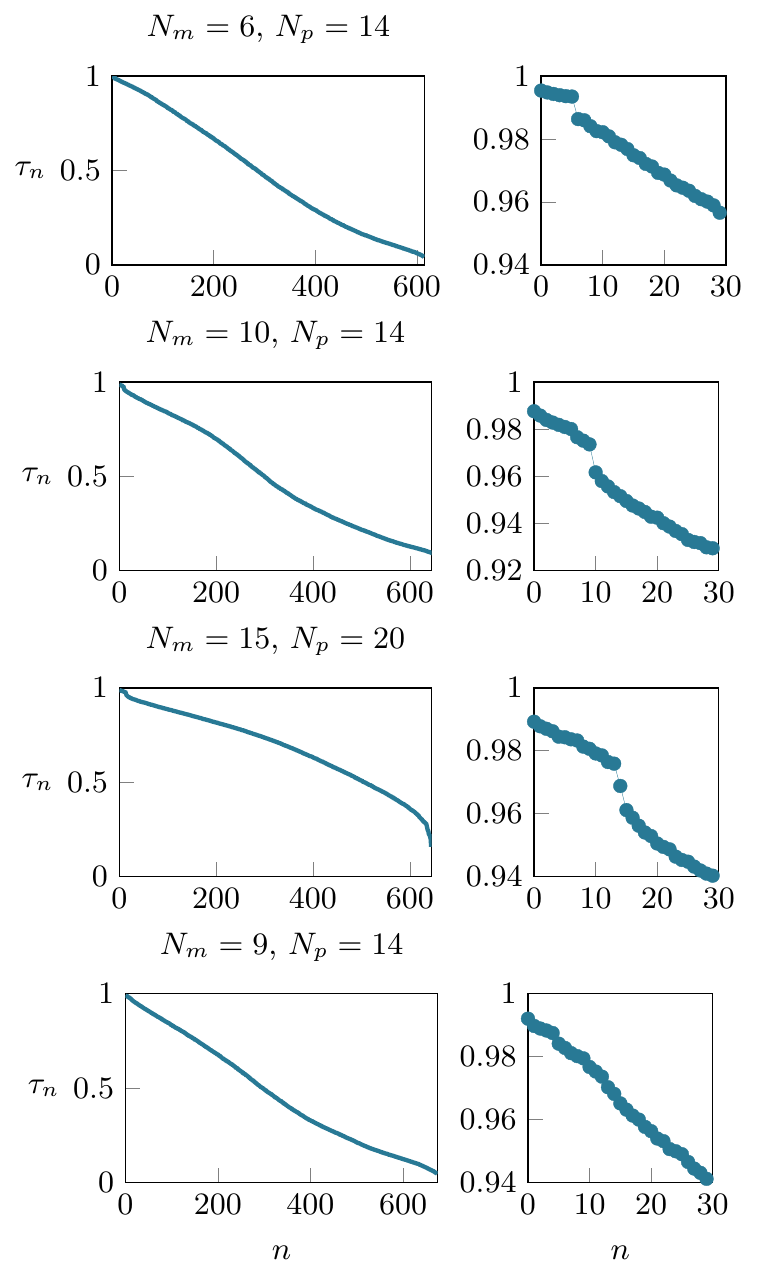}
\caption{Singular values for different MPLC systems with $N_m = 6, 10, 15, 9$ and $N_p = 14, 14, 20, 14$. On the left we plot all the singular values while on the right we only plot the $30$ largest ones.
The systems corresponding to the first three rows were designed to manipulate the guided modes of three different optical fibres, while the one in the last row was designed to shape Laguerre-Gaussian modes.}
\label{fig:singularvalues}
\end{figure}

Several important properties of a scattering medium, e.g. its total transmissivity, can be obtained from the singular value decomposition (SVD) of its transmission matrix $t$ \cite{Rotter_Gigan}.
The latter is defined as,
\begin{equation}
t = U D V^\dagger,
\end{equation}
where $U$ and $V$ are unitary matrices of dimensions $Q\times Q$ and $P\times P$, while $D = \rm{diag} (\tau_1, \cdots, \tau_P)$ is a $Q\times P$ diagonal matrix containing the $P$ singular values $\tau$ of $t$.
The singular values can be calculated as the square roots of the standard eigenvalues $T$ of the $P\times P$ Hermitian matrix $t^\dagger t$, i.e. $\tau = \sqrt{T}$ \cite{Rotter_Gigan}.

For an ideal MPLC system, the first $N_m$ singular values are exactly equal to one, i.e.
\begin{equation}
D= \rm{diag} (\overbrace{1, \cdots, 1}^{N_m}, \overbrace{\vphantom{1}\tau_{N_{m+1}}, \cdots, \tau_P}^{P-N_m}),
\end{equation}
while the first $N_m$ left and right singular vectors, contained in $U$ and $V$ respectively, are given by orthogonal linear combinations of the output $\left(\{u_i\}_{1\leq i \leq N_m}\right)$ and input $\left(\{v_i\}_{1\leq i \leq N_m}\right)$ design modes.

In practical implementations, a combination of suboptimal design and losses induces a deviation of the first $N_m$ singular values from unity. 
For the same reasons, the first $N_m$ singular vectors of a realistic device will acquire finite contributions from modes different from the design ones.
These deviations can therefore be used to evaluate the quality of the design of an MPLC system.
On the other hand, the other $P - N_m$ singular values and singular vectors describe the transmission properties of the device outside of the design subspace.

In Fig. \ref{fig:singularvalues}, we plot the singular values of four different MPLC systems, which are distinguished by the number of shaped modes ($N_m = 6, 10, 15, 9$), as well as the number of phase plates ($N_p = 14, 14, 20, 14$) used to construct them. 
All four systems map spatially separated Gaussian spots in the input plane (the modes $\{v_i\}_{1\leq i \leq N_m}$) into a set of orthogonal co-propagating modes $\{u_i\}_{1\leq i \leq N_m}$.
For the systems corresponding to the first three rows in Fig. \ref{fig:singularvalues}, the output modes $\{u_i\}_{1\leq i \leq N_m}$ are different numbers of optical fibers' modes, while for the one corresponding to the bottom row they are Laguerre-Gauss (LG) modes.

In Fig. \ref{fig:singularvalues}, we observe that the first $N_m$ singular values stand out from the others: a gap appears.
This feature, which is not surprising since these systems are optimized to shape and transmit preferentially $N_m$ modes, can be used to compare the performances (in terms of transmission losses) of different MPLC devices.
In fact, the amplitude of the gap depends on the design transformation, as we can notice in the bottom row of Fig. \ref{fig:singularvalues}, where the difference between the first $N_m$ singular values and the rest of them is practically unnoticeable.
This suggests that the conversion to LG-modes is less efficient than the transformations to fiber-modes performed by the other three devices considered.
We will confirm this fact in the following, with a detailed singular modes analysis.
Moreover, it should be noted that the amplitude of the gap is dependent on the optimization algorithm used to design the MPLC systems. 
The devices associated with the singular-value distributions in Fig. \ref{fig:singularvalues} are all multiplexing devices constructed for telecommunication purposes, and therefore optimized to reduce the crosstalk among the output modes.
In a different context, an optimization algorithm which focuses on reducing losses could be used to enhance this gap between the first $N_m$ singular values and the others.

\begin{figure*}
\begin{tikzpicture}
\node at (0,0) {\includegraphics[width=0.98\textwidth]{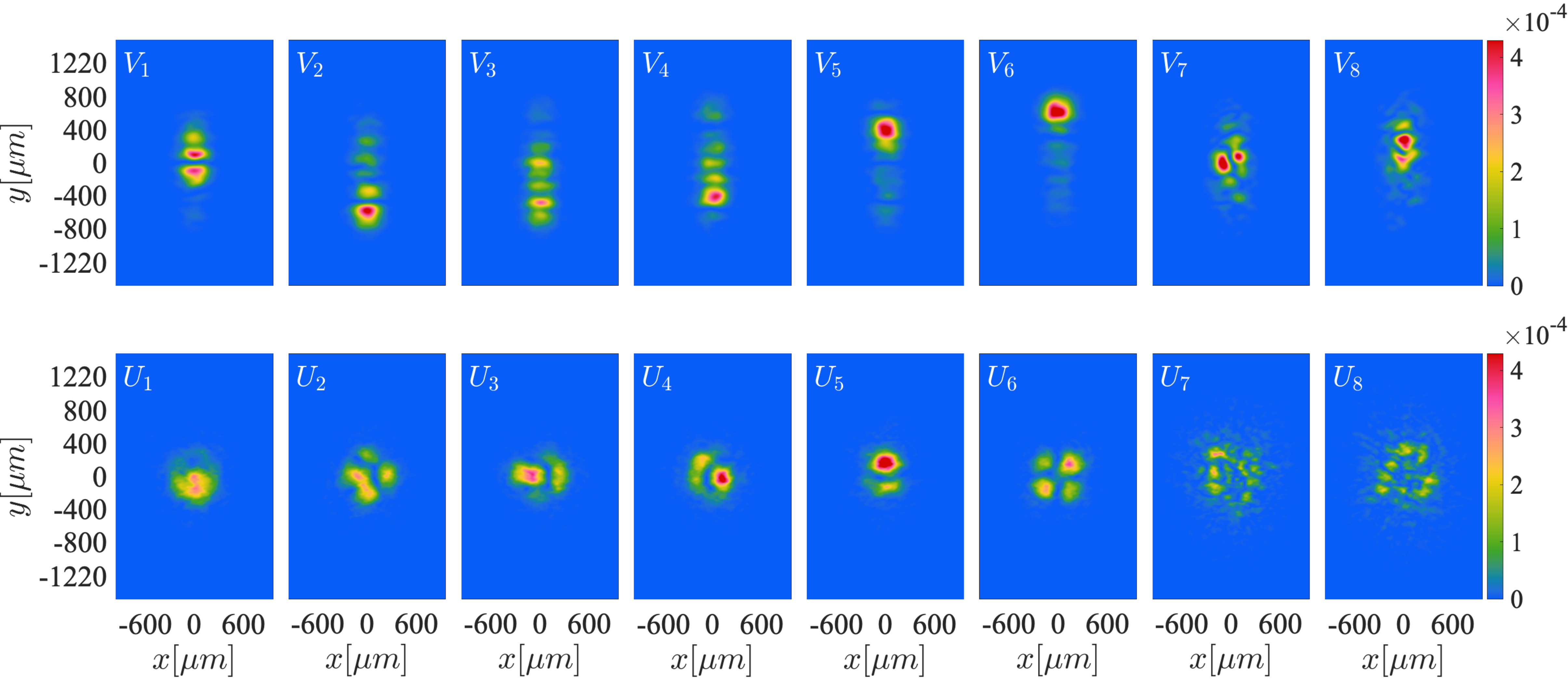}};
\draw[ultra thick,black!30!green] (-7.53,-2.95) -- (-7.53,3.42) --  (4.025,3.42) -- (4.025,-2.95) -- cycle;
\draw[ultra thick,black!20!red] (7.83,-2.95) -- (7.83,3.42) -- (4.075,3.42) -- (4.075,-2.95) -- cycle;
\end{tikzpicture}
\caption{Left (top row) and right (bottom row) singular modes $\{U_i\}_{1\leq i \leq Q}$ and $\{V_i\}_{1\leq i \leq P}$ of the MPLC system with $N_m=6$ and $N_P=14$, whose singular values are showed in the top panel of Fig. \ref{fig:singularvalues}. 
The axes of the numerical grid have the same size as the actual phase plates of the device and are centered with respect to them. 
The modes in the green rectangle correspond to the $N_m=6$ largest singular values, and are dominated by linear superpositions of the design modes. 
On the other hand, the other singular modes (red rectangle) look like speckle patterns.}
\label{Fig:singular_modes}
\end{figure*}
Let us now have a look at the left and right singular modes $\{U_i\}_{1\leq i \leq Q}$ and $\{V_i\}_{1\leq i \leq P}$ of the transmission matrix $t$, which are obtained from the unitary matrices $U$ and $V$ according to
\begin{align}
U_i = \sum_{k=1}^Q U_{ik}\psi_k,\\
V_i = \sum_{k=1}^P V_{ik}\phi_k,
\end{align}
In all analysed cases, the singular modes corresponding to the largest $N_m$ singular values --- i.e. the most efficiently transmitted singular modes --- are close to linear superpositions of the design modes. 
This is evident in Fig. \ref{Fig:singular_modes} where some singular modes of an MPLC system designed to map $N_m=6$ Gaussian spots aligned along the $y-$direction into an equal number of modes of an optical fiber are showed. 
The design modes of this particular MPLC device are those shown in Fig. \ref{fig:modelisationMPLC}, while its singular values are plotted in the top panel of Fig. \ref{fig:singularvalues}.
We can clearly see (top row of Fig. \ref{Fig:singular_modes}) that the right singular modes corresponding to the $6$ largest singular values ($V_{1} -V_{6}$) of this MPLC system correspond to a linear superposition of the Gaussian spots in the top row of Fig. \ref{fig:modelisationMPLC}. 
In a similar fashion the modes, $U_{1} - U_{6}$ (bottom row of Fig. \ref{Fig:singular_modes}) correspond to linear superposition of the design fiber modes (bottom row of Fig. \ref{fig:modelisationMPLC}).
To make this observation more quantitative, let us consider which fraction of the power of the design output modes $\{u_i\}_{1\leq i\leq N_m}$ is contained into the subspace spanned by the best transmitting left singular modes $\{U_j\}_{1\leq j\leq N_m}$. Such fraction can be computed as
\begin{equation}
p_i = \sum_{j=1}^{N_m}\left|\int u_i^*(x,y) U_j (x,y) d x d y\right|^2.
\label{p}
\end{equation}
The quantity \eqref{p} is reported in table \ref{tab:overlap} for all the design modes $u_i$ of the four MPLC systems whose singular values are plotted in Fig. \ref{fig:singularvalues}.
\begin{table*}
\begin{ruledtabular}
	\centering
\begin{tabular*}{\linewidth}{@{\extracolsep{\fill}}cc|ccccccccccccccc}
$N_m$ & $N_p$	& 1 & 2 & 3 & 4 & 5 & 6 & 7 & 8 & 9 & 10 & 11 & 12 & 13 & 14 & 15\\
		\colrule
6 & 14	& 0.940 & 0.938 & 0.953 & 0.939 & 0.916 & 0.949 & & & & & & & & & \\ 9 & 14 & 0.413 & 0.682 & 0.835 & 0.864 & 0.849 & 0.863 & 0.786 & 0.738 & 0.802 & & & & & & \\
10 & 14 &		0.887 & 0.897 & 0.909 & 0.845 & 0.883 & 0.863 & 0.838 & 0.861 & 0.874 & 0.902 & & & & & \\
15 & 20 &		0.838 & 0.877 & 0.898 & 0.927 & 0.902 & 0.907 & 0.898 & 0.899 & 0.882 & 
0.857 & 0.874 & 0.900 & 0.900 & 0.865 & 0.824 
\end{tabular*}
\end{ruledtabular}
\caption{Power fraction of the design output modes $\{u_i\}_{1\leq i\leq N_m}$ contained into the subspace spanned by the best transmitting left singular modes $\{U_j\}_{1\leq j\leq N_m}$ for the four MPLC systems whose singular values are plotted in Fig. \ref{fig:singularvalues} (see \eqref{p}). The first two columns report the number of shaped modes $N_m$, and the number of phase plates $N_P$ characterizing the MPLC devices, while the following columns contain the values of $p_i$.}
\label{tab:overlap}
\end{table*}
The large values of $p_i$ reported in table \ref{tab:overlap} confirm that the design transformation is almost completely represented in the subspace of the most efficiently transmitted modes. 
This means that the studied devices make a very efficient use of the high-dimensional mode space at their disposal, relying on almost exactly as much degrees of freedom as needed to realize the design transformation.
One should however notice that the values of $p_i$ for the device corresponding to the second row of table \ref{tab:overlap} are lower (significantly lower in the case of $p_1$ and $p_2$) than those of the other devices.
We remind the reader, that this device's output modes $u_i$ are LG modes, and that its singular value distribution doesn't feature a gap (see bottom row of Fig. \ref{fig:singularvalues}). 
The main difference between the LG modes, and the fiber modes at the output of the other MPLC systems considered in Fig. \ref{fig:singularvalues} and table \ref{tab:overlap} is that the width of the LG modes quickly increases with their mode order.
Therefore, it is, arguably, harder for the MPLC system to widen the input modes to the proper size within a limited number of reflections on phase plates.
This results in the need of a larger set of singular modes to properly represent the design mode transformation.
Or, in other words, this MPLC system is making a slightly less efficient use of the resources at its disposal to realize the design transformation.

The above discussion is an example of how SVD can provide information about the quality of MPLC transformations and the optimality of their design.
Let us now move our attention to the singular modes $U_i$ and $V_i$ with $i>N_m$.
In general, these modes bear no resemblance with the design modes, and, especially in the output plane, look like speckle patterns (see $U_{7}, U_{8}$, and $V_{7}, V_{8}$ in Fig. \ref{fig:singularvalues}).
This observation, together with the high dimensionality and complexity of the MPLC transformation, suggests that an MPLC system essentially behaves as a chaotic cavity. 
In the following section, we will build on this behaviour to derive an analytical model for the transmission properties of MPLC devices.

\section{Analytical theory}
\label{Sec:FRM}

Let us therefore consider an MPLC device as a scattering system. 
As such, it can be described by a $2N \times 2N$ scattering matrix
\begin{equation}
S =
\begin{pmatrix}
r_0 & t'_0 \\
t_0 & r'_0
\end{pmatrix},
\label{S}
\end{equation}
with $N$ the total number of spatial modes supported by the system.
Accordingly, $r_0$ ($r_0^\prime$) and $t_0$ ($t_0^\prime$) are blocks of size $N\times N$ and determine the amplitudes of the modes which, incoming from the left (right) in Fig. \ref{fig:modelisationMPLC}, are, respectively, reflected and transmitted by the MPLC system.
In real devices, diffraction causes a portion of the injected light to go beyond the physical extent of the phase plates. 
This effect limits the number of spatial modes that can be controlled by a particular system.
Therefore, in practice, one does not have access to the full transmission matrix $t_0$, but rather to a submatrix $t$ which is obtained by {\it filtering} $t_0$, i.e. by removing $N-N_1$ columns and $N-N_2$ rows of $t_0$.
Here, $N_1$ and $N_2$ represent the number of spatial modes that can be controlled in the input and output planes respectively. 
We will refer to these modes as the {\it controllable modes}. They are determined by physical constraints of the system, and are, in general, unknown. 
The reader should not confuse them with the modes the system is designed to shape, nor with the modes $\phi_i$ and $\psi_i$ we used in Sec. \ref{Sec:transmission} to obtain an accurate numerical representation of $t$.

Given that a perfect scattering system does not have losses, its scattering matrix $S$ is by definition unitary: $S^{\dagger}S = 1$.
In addition, given the high dimensionality and complexity of an MPLC system, 
we will treat its scattering matrix as a random matrix, similarly to what is done in condensed matter physics for characterizing transport in quantum dots or metal wires \cite{Bennaker_1997}.
When increasing the number $N_m$ of modes to be shaped, in order to resolve finer spatial structures, the patterns to be impressed on the phase plates get finer, and, thereby, look more random.
Therefore, we expect the random matrix theory approach to become valid for large values of $N_m$.

Considering the transmission matrix $S$ as random allows us to use filtered random matrix (FRM) theory to derive the probability distribution $\rho_{t^\dagger t}(T)$ of the eigenvalues $T$ of the matrix $t^\dagger t$, which is related to the probability distribution $\rho_t(\tau)$ of the singular values $\tau$ of the transmission matrix $t$ according to 
$\rho_t(\tau) = 2\tau \rho_{t^\dagger t}(\tau^2)$ (see Sec. \ref{Sec:transmission}).

\subsection{Filtered random matrix model}
Let us notice that the extraction of the transmission matrix $t_0$ from the scattering matrix $S$ can be considered as a filtering where $N$ rows and $N$ columns are removed.
Accordingly, $t$ is obtained from two successive filtering, a first one to extract $t_0$ from $S$, and a second to extract $t$ from $t_0$. 
In the following, we recall the general FRM formalism, and then apply it twice to derive $\rho_{t^\dagger t}( T)$.

Given an $M \times M$ random matrix $A$, the eigenvalue density of the Hermitian matrix $A^\dagger A$ is given by
\begin{equation}
\rho_{A^\dagger A} (T) = -\frac{1}{\pi}\lim_{\epsilon\to 0} \rm{Im} \left[ g_{A^\dagger A}(T + i \epsilon)\right],
\label{rhoA}
\end{equation}
where we have introduced the resolvent 
\begin{equation}
g_{A^\dagger A}(z) = \frac{1}{M}\left \langle \rm{Tr} \frac{1}{z - A^\dagger A} \right\rangle,
\label{gA}
\end{equation}
with $\langle \cdots \rangle$ denoting the ensemble average \cite{tulino2004}.
Let us now consider the {\it filtered} random matrix
\begin{equation}
\tilde{A} = P_2 A P_1,
\end{equation}
with $P_1$ and $P_2$ two matrices of sizes $M\times M_1$ and $M_2\times M$, respectively, that eliminate $M-M_1$ columns and $M-M_2$ rows of $A$.
The resolvent $g_{\tilde{A}^\dagger \tilde{A}}(z)$ of $\tilde{A}^\dagger \tilde{A}$ is connected to $g_{A^\dagger A}(z)$ by the FRM equation \cite{Goetschy13}
\begin{equation}
N(z) g_{A^\dagger A}\left(\frac{N^2(z)}{D(z)}\right) = D(z),
\label{FRMeq}
\end{equation}
where $N(z)$ and $D(z)$ are defined according to 
\begin{subequations}
\begin{align}
N(z) &= z m_1 g_{\tilde{A}^\dagger \tilde{A}}(z) + 1 - m_1,\\
D(z) &= m_1g_{\tilde{A}^\dagger \tilde{A}}(z)\left[ zm_1g_{\tilde{A}^\dagger \tilde{A}}(z) + m_2 -m_1\right],
\end{align}
\end{subequations}
with the filtering parameters $m_1 = M_1/M$ and $m_2 =M_2/M$.

Let us now apply the FRM equation \eqref{FRMeq} to derive the resolvent of $t_0^\dagger t_0$ from the one of $S^\dagger S$.
By using the unitarity of $S$ and Eq. \eqref{gA}, we have $g_{S^\dagger S} (z) = 1/(z-1)$, which, together with Eq. \eqref{FRMeq} with filtering parameters $m_1 = m_2 = 1/2$, gives us
\begin{equation}
g_{t_0^\dagger t_0}(z) = \frac{1}{\sqrt{z(z-1)}}.
\label{gt0}
\end{equation}
The eigenvalue density associated with the resolvent \eqref{gt0} [see Eq. \eqref{rhoA}] corresponds to the well-known {\it bimodal distribution} associated to chaotic cavities \cite{Bennaker_1997}
\begin{equation}
\rho_{t_0^\dagger t_0}(T) = \frac{1}{\pi}\frac{1}{\sqrt{T(1-T)}}.
\end{equation}
We now apply Eq. \eqref{FRMeq} once more, this time with filtering parameters $m_1 =N_1/N$ and $m_2 = N_2/N$ with $N_1$ and $N_2$ the number of controllable modes, to obtain the resolvent of $t^\dagger t$ 
\begin{align}\label{gt}
g_{t^\dagger t}\left(z\right) &= \frac{1}{2 m_1 z \left(1-z\right)}\biggl( m_1 - m_2 + 2\left(1-m_1\right)z  \biggr.  \\
& \left. -\left[\left(m_1 - m_2\right)^2+4z^2 - 4\left(m_1+ m_2 -m_1 m_2 \right)z\right]^{1/2}\right).\nonumber
\end{align}
Finally, by using Eq. \eqref{rhoA}, we obtain the transmission-eigenvalue density
\begin{align}
\rho_{t^\dagger t} (T) &= \frac{1}{\pi} \frac{\sqrt{\left(T^+-T\right)\left(T - T^-\right)}}{m_1 T \left(1 - T\right)} \nonumber \\ &+ \text{max}\left(1-\frac{m_2}{m_1}, 0\right)\delta\left(T\right)
\label{rhot}
\end{align}
with
\begin{equation}
T^{\pm} = \frac{m_1 + m_2 - m_1 m_2 \pm \sqrt{m_1 m_2 \left(2-m_1\right)\left(2 - m_2\right)}}{2}.
\end{equation} 

\subsection{Comparison with numerical results}
\begin{figure}[tb]
\includegraphics[width=0.9\columnwidth]{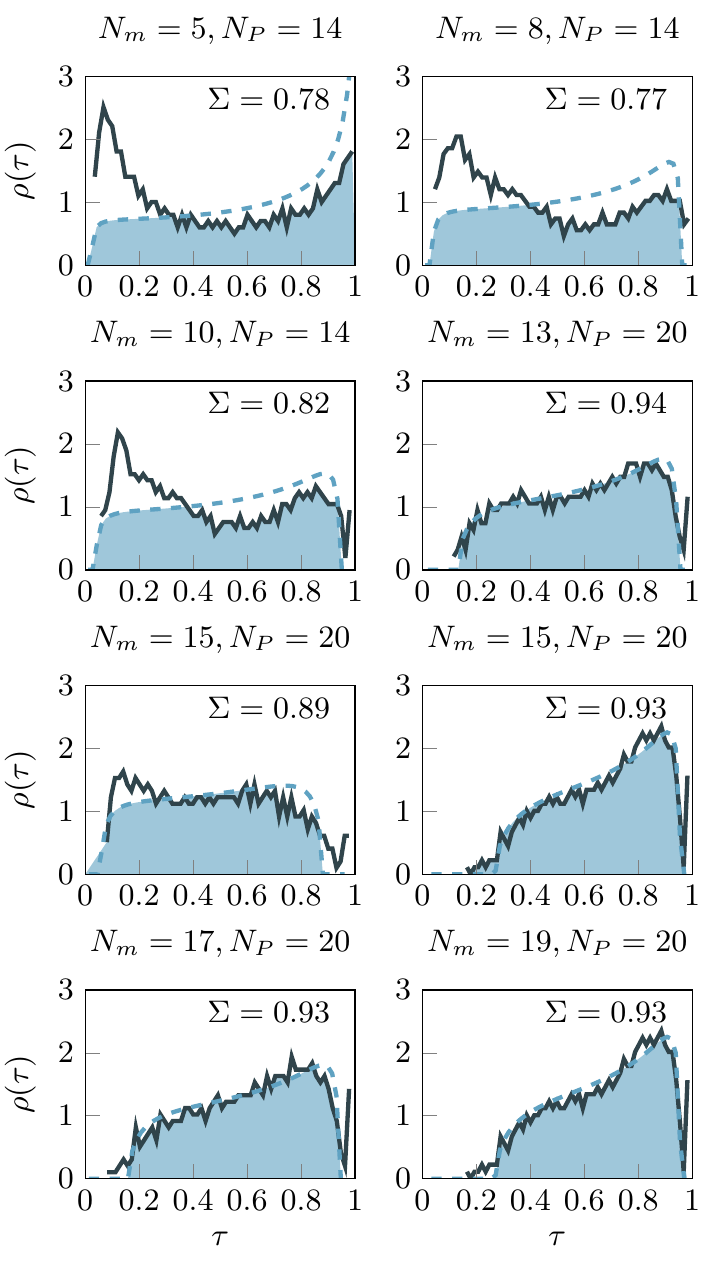}
\caption{Full lines: singular value distributions obtained from the numerical simulation of different MPLC systems with $N_m = 5, 8, 10, 13, 15, 17, 19$, and $N_P = 14, 20$. Dashed lines: filtered random matrix model obtained from Eq. \eqref{rhot} that best fits the data. 
The corresponding fit coefficients are presented in table \ref{Tab:fit_parameters}.
The shaded area represents the similarity function $\Sigma$ whose value is given in the upper right corner.}
\label{fig:model0_a}
\end{figure}

Let us now compare the prediction of the FRM theory with the numerical data obtained from the MPLC devices defined in Sec. \ref{Sec:definition}.
In general, the results of random matrix theory are valid when an average over several elements of an ensemble is considered.
However, for large enough matrices, a {\it self-averaging or ergodicity} argument can be invoked, i.e we can assume that a single matrix is sufficient to represent the whole ensemble \cite{tulino2004}.
The transmission matrices $t$ computed in section \ref{Sec:transmission} satisfy this self-averaging argument. 
Accordingly, we fit the probability distribution of the singular values, $\rho_{t}(\tau)$, extracted from the numerical data presented in Sec. \ref{Sec:SVD} to those obtained from Eq. \eqref{rhot}. 

The fits were performed by optimizing the parameters $m_1$ and $m_2$ in order to maximize the similarity function $\Sigma$ defined as the area under the point-by-point minimum of the data and the model curves (shaded area in Fig. \ref{fig:model0_a}). 
Given that the singular value distribution $\rho_t(\tau)$ is normalized to unity, the similarity function $\Sigma\in[0,1]$.
The curves resulting from this fitting procedure for eight different MPLC systems are presented in Fig. \ref{fig:model0_a}. 
The corresponding fitting parameters $m_1$ and $m_2$ are listed in table \ref{Tab:fit_parameters}.

Fig. \ref{fig:model0_a} shows that, for MPLC systems designed to shape a low number $N_m$ of modes by using a low number $N_P$ of phase plates, the distribution $\rho_t(\tau)$ presents a peak at low singular values which correspond to singular modes with low transmittance.
Such a peak is not well fitted by our analytical model. 
On the other hand, when increasing the number $N_m$ of shaped modes as well as the number $N_P$ of phase plates, the numerical singular value distributions are very well reproduced by FRM model.
This behaviour fits with our intuition that the assumption for the scattering matrix $S$ being random is justified only for systems designed to shape a large number of modes.
\begin{table}[t]
\begin{ruledtabular}
\centering
\begin{tabular}{c| cccccccc}
$N_m$	& 5 & 8 & 10 & 13 & 15 & 15 & 17 & 19\\
\colrule
$m_1$	& 0.87 & 0.68 & 0.64 & 0.56 & 0.47 & 0.46 & 0.52 & 0.49 \\
$m_2$ & 0.91 & 0.75 & 0.71 & 0.81 & 0.58 & 0.98 & 0.82 & 0.8
\end{tabular}
\end{ruledtabular}
\caption{Fitting parameters $m_1$ and $m_2$ corresponding to the curves presented in Fig. \ref{fig:model0_a} }
\label{Tab:fit_parameters}
\end{table}

The third row of Fig. \ref{fig:model0_a} shows the singular value distributions of two MPLC systems that use the same number $N_P=20$ of phase plates to transform the same number $N_m =15$ of separated Gaussian spots into modes from two different mode families (e.g. the guided modes of two different optical fibers).
For both systems, our analytical model fits quite well the numerical data, but with different values of the fitting parameters (see in particular $m_2$ in table \ref{Tab:fit_parameters}).
We therefore conclude that, when the number of shaped modes is large, the overall transmission properties of an MPLC system are those of a random scattering system with a limited number of controllable modes. 
However, the different values of $m_1$ and $m_2$ tell us that the exact number of controllable modes can be strongly influenced by the spatial profile of the modes that the system is designed to shape.

Moreover, by looking at the fit parameters in table \ref{Tab:fit_parameters}, we note that $m_1$ tends to get smaller when the number of shaped modes $N_m$ increases.
This is probably due to the fact that, in order to manipulate higher-order spatial modes, it is necessary to enlarge the area of the patterns inscribed onto the phase plates.
As a consequence, diffraction pushes more and more light beyond the edges of the phase plates and the fraction of controlled channels, as quantified by $m_1$ and $m_2$, decreases.

\section{Conclusion}
\label{Sec:conclusion}
In this work, we presented a complete characterization of the transmission properties of MPLC systems, and, for the first time, we investigated the behaviour of these systems outside the subspace of modes that they are designed to shape.

Our analysis shows how the singular value decomposition of the MPLC systems' transmission matrices can be a powerful tool to quantify the performances of these devices.
In particular, we studied the overlap between the subspace spanned by the singular modes corresponding to the largest $N_m$ singular values and the one spanned by the design modes. 
This quantity provides a clear indication on how efficiently an MPLC system can use the high-dimensional resources at its disposal to realize the design transformation.

Together with the numerical results, we introduced a filtered random matrix analytical model, which very well captures the probability distribution of the singular values of systems designed to shape a large number of modes.
Such a good agreement with our analytical model suggests that in these cases an MPLC system behaves like a chaotic cavity or random scattering medium with only a limited number of controllable modes.

The results of our analysis provide new elements to evaluate and predict the performances of MPLC systems.
For example, we could predict the amplitude of the largest singular values from our analytical model, and use the fact that these singular values are associated with the design modes to put a bound on the total transmittance of an MPLC transformation.
Moreover, our study of the transmission properties outside of the design-mode subspace brings to light new parameters that could be optimized in the construction of MPLC devices.
For instance, one could enhance the gap between the largest $N_m$ singular values and the others. 
Doing so, one would increase the losses experienced by injecting into the MPLC system modes outside of the design subspace, e.g. by misaligning the system.
The result would be a device which could be easily aligned simply by monitoring the transmitted power. 

On a larger scope, these findings forge a connection between highly tuned optical technology and the physics of complex media. 
As such, we are exploring the tension between, on the one hand, control and design, and, on the other hand, complexity. 
Our results impose new fundamental questions, e.g., about the point at which the system transitions towards the physics of a random optical medium (as shown in Fig. 4). 
More microscopic models will be needed to understand how the statistical features of MPLC ultimately sum up to reproduce that statistics of a random matrix model, and to understand the role of different design parameters in this process. 
Then, ultimately, we may hope to exploit the chaotic statistics that manifests in the system to improve the design of such optical technology. 

\begin{acknowledgments}
The authors thank Guillaume Labroille from Cailabs for providing access to the MPLC systems' data and for helpful discussions.
\end{acknowledgments}

\bibliography{apssamp}

\providecommand{\noopsort}[1]{}\providecommand{\singleletter}[1]{#1}%
\begin{thebibliography}{22}%
\makeatletter
\providecommand \@ifxundefined [1]{%
 \@ifx{#1\undefined}
}%
\providecommand \@ifnum [1]{%
 \ifnum #1\expandafter \@firstoftwo
 \else \expandafter \@secondoftwo
 \fi
}%
\providecommand \@ifx [1]{%
 \ifx #1\expandafter \@firstoftwo
 \else \expandafter \@secondoftwo
 \fi
}%
\providecommand \natexlab [1]{#1}%
\providecommand \enquote  [1]{``#1''}%
\providecommand \bibnamefont  [1]{#1}%
\providecommand \bibfnamefont [1]{#1}%
\providecommand \citenamefont [1]{#1}%
\providecommand \href@noop [0]{\@secondoftwo}%
\providecommand \href [0]{\begingroup \@sanitize@url \@href}%
\providecommand \@href[1]{\@@startlink{#1}\@@href}%
\providecommand \@@href[1]{\endgroup#1\@@endlink}%
\providecommand \@sanitize@url [0]{\catcode `\\12\catcode `\$12\catcode
  `\&12\catcode `\#12\catcode `\^12\catcode `\_12\catcode `\%12\relax}%
\providecommand \@@startlink[1]{}%
\providecommand \@@endlink[0]{}%
\providecommand \url  [0]{\begingroup\@sanitize@url \@url }%
\providecommand \@url [1]{\endgroup\@href {#1}{\urlprefix }}%
\providecommand \urlprefix  [0]{URL }%
\providecommand \Eprint [0]{\href }%
\providecommand \doibase [0]{http://dx.doi.org/}%
\providecommand \selectlanguage [0]{\@gobble}%
\providecommand \bibinfo  [0]{\@secondoftwo}%
\providecommand \bibfield  [0]{\@secondoftwo}%
\providecommand \translation [1]{[#1]}%
\providecommand \BibitemOpen [0]{}%
\providecommand \bibitemStop [0]{}%
\providecommand \bibitemNoStop [0]{.\EOS\space}%
\providecommand \EOS [0]{\spacefactor3000\relax}%
\providecommand \BibitemShut  [1]{\csname bibitem#1\endcsname}%
\let\auto@bib@innerbib\@empty
\bibitem [{\citenamefont {Vellekoop}(2015)}]{Vellekoop:15}%
  \BibitemOpen
  \bibfield  {author} {\bibinfo {author} {\bibfnamefont {I.~M.}\ \bibnamefont
  {Vellekoop}},\ }\href {\doibase 10.1364/OE.23.012189} {\bibfield  {journal}
  {\bibinfo  {journal} {Opt. Express}\ }\textbf {\bibinfo {volume} {23}},\
  \bibinfo {pages} {12189} (\bibinfo {year} {2015})}\BibitemShut {NoStop}%
\bibitem [{\citenamefont {Yu}\ \emph {et~al.}(2015)\citenamefont {Yu},
  \citenamefont {Park}, \citenamefont {Lee}, \citenamefont {Yoon},
  \citenamefont {Kim}, \citenamefont {Lee},\ and\ \citenamefont
  {Park}}]{YU2015632}%
  \BibitemOpen
  \bibfield  {author} {\bibinfo {author} {\bibfnamefont {H.}~\bibnamefont
  {Yu}}, \bibinfo {author} {\bibfnamefont {J.}~\bibnamefont {Park}}, \bibinfo
  {author} {\bibfnamefont {K.}~\bibnamefont {Lee}}, \bibinfo {author}
  {\bibfnamefont {J.}~\bibnamefont {Yoon}}, \bibinfo {author} {\bibfnamefont
  {K.}~\bibnamefont {Kim}}, \bibinfo {author} {\bibfnamefont {S.}~\bibnamefont
  {Lee}}, \ and\ \bibinfo {author} {\bibfnamefont {Y.}~\bibnamefont {Park}},\
  }\href {\doibase https://doi.org/10.1016/j.cap.2015.02.015} {\bibfield
  {journal} {\bibinfo  {journal} {Current Applied Physics}\ }\textbf {\bibinfo
  {volume} {15}},\ \bibinfo {pages} {632 } (\bibinfo {year}
  {2015})}\BibitemShut {NoStop}%
\bibitem [{\citenamefont {Schwartz}\ \emph {et~al.}(2009)\citenamefont
  {Schwartz}, \citenamefont {V{\'e}drenne}, \citenamefont {Michau},
  \citenamefont {Velluet},\ and\ \citenamefont {Chazallet}}]{schwartz2009}%
  \BibitemOpen
  \bibfield  {author} {\bibinfo {author} {\bibfnamefont {N.~H.}\ \bibnamefont
  {Schwartz}}, \bibinfo {author} {\bibfnamefont {N.}~\bibnamefont
  {V{\'e}drenne}}, \bibinfo {author} {\bibfnamefont {V.}~\bibnamefont
  {Michau}}, \bibinfo {author} {\bibfnamefont {M.-T.}\ \bibnamefont {Velluet}},
  \ and\ \bibinfo {author} {\bibfnamefont {F.}~\bibnamefont {Chazallet}},\
  }\href@noop {} {\bibfield  {journal} {\bibinfo  {journal} {Atmospheric
  Propagation of Electromagnetic Waves III}\ }\textbf {\bibinfo {volume}
  {7200}},\ \bibinfo {pages} {72000J} (\bibinfo {year} {2009})}\BibitemShut
  {NoStop}%
\bibitem [{\citenamefont {Sorelli}\ \emph {et~al.}(2019)\citenamefont
  {Sorelli}, \citenamefont {Leonhard}, \citenamefont {Shatokhin}, \citenamefont
  {Reinlein},\ and\ \citenamefont {Buchleitner}}]{Sorelli_2019}%
  \BibitemOpen
  \bibfield  {author} {\bibinfo {author} {\bibfnamefont {G.}~\bibnamefont
  {Sorelli}}, \bibinfo {author} {\bibfnamefont {N.}~\bibnamefont {Leonhard}},
  \bibinfo {author} {\bibfnamefont {V.~N.}\ \bibnamefont {Shatokhin}}, \bibinfo
  {author} {\bibfnamefont {C.}~\bibnamefont {Reinlein}}, \ and\ \bibinfo
  {author} {\bibfnamefont {A.}~\bibnamefont {Buchleitner}},\ }\href {\doibase
  10.1088/1367-2630/ab006e} {\bibfield  {journal} {\bibinfo  {journal} {New
  Journal of Physics}\ }\textbf {\bibinfo {volume} {21}},\ \bibinfo {pages}
  {023003} (\bibinfo {year} {2019})}\BibitemShut {NoStop}%
\bibitem [{\citenamefont {Defienne}\ \emph {et~al.}(2016)\citenamefont
  {Defienne}, \citenamefont {Barbieri}, \citenamefont {Walmsley}, \citenamefont
  {Smith},\ and\ \citenamefont {Gigan}}]{Defienne:2016}%
  \BibitemOpen
  \bibfield  {author} {\bibinfo {author} {\bibfnamefont {H.}~\bibnamefont
  {Defienne}}, \bibinfo {author} {\bibfnamefont {M.}~\bibnamefont {Barbieri}},
  \bibinfo {author} {\bibfnamefont {I.~A.}\ \bibnamefont {Walmsley}}, \bibinfo
  {author} {\bibfnamefont {B.~J.}\ \bibnamefont {Smith}}, \ and\ \bibinfo
  {author} {\bibfnamefont {S.}~\bibnamefont {Gigan}},\ }\href
  {https://advances.sciencemag.org/content/2/1/e1501054} {\bibfield  {journal}
  {\bibinfo  {journal} {Science Advances}\ }\textbf {\bibinfo {volume} {2}},\
  \bibinfo {pages} {e1501054} (\bibinfo {year} {2016})}\BibitemShut {NoStop}%
\bibitem [{\citenamefont {Tyson}(2010)}]{tyson}%
  \BibitemOpen
  \bibfield  {author} {\bibinfo {author} {\bibfnamefont {R.}~\bibnamefont
  {Tyson}},\ }\href@noop {} {\emph {\bibinfo {title} {Principles of Adaptive
  Optics}}}\ (\bibinfo  {publisher} {CRC Press},\ \bibinfo {address} {Boca
  Raton},\ \bibinfo {year} {2010})\BibitemShut {NoStop}%
\bibitem [{\citenamefont {Rotter}\ and\ \citenamefont
  {Gigan}(2017)}]{Rotter_Gigan}%
  \BibitemOpen
  \bibfield  {author} {\bibinfo {author} {\bibfnamefont {S.}~\bibnamefont
  {Rotter}}\ and\ \bibinfo {author} {\bibfnamefont {S.}~\bibnamefont {Gigan}},\
  }\href {\doibase 10.1103/RevModPhys.89.015005} {\bibfield  {journal}
  {\bibinfo  {journal} {Rev. Mod. Phys.}\ }\textbf {\bibinfo {volume} {89}},\
  \bibinfo {pages} {015005} (\bibinfo {year} {2017})}\BibitemShut {NoStop}%
\bibitem [{\citenamefont {Matth\`{e}s}\ \emph {et~al.}(2019)\citenamefont
  {Matth\`{e}s}, \citenamefont {del Hougne}, \citenamefont {de~Rosny},
  \citenamefont {Lerosey},\ and\ \citenamefont {Popoff}}]{Matthess:2019}%
  \BibitemOpen
  \bibfield  {author} {\bibinfo {author} {\bibfnamefont {M.~W.}\ \bibnamefont
  {Matth\`{e}s}}, \bibinfo {author} {\bibfnamefont {P.}~\bibnamefont {del
  Hougne}}, \bibinfo {author} {\bibfnamefont {J.}~\bibnamefont {de~Rosny}},
  \bibinfo {author} {\bibfnamefont {G.}~\bibnamefont {Lerosey}}, \ and\
  \bibinfo {author} {\bibfnamefont {S.~M.}\ \bibnamefont {Popoff}},\
  }\href@noop {} {\bibfield  {journal} {\bibinfo  {journal} {Optica}\ }\textbf
  {\bibinfo {volume} {6}},\ \bibinfo {pages} {465} (\bibinfo {year}
  {2019})}\BibitemShut {NoStop}%
\bibitem [{\citenamefont {Su}\ \emph {et~al.}(2018)\citenamefont {Su},
  \citenamefont {Piggott}, \citenamefont {Sapra}, \citenamefont {Petykiewicz},\
  and\ \citenamefont {Vuckovic}}]{su2018inverse}%
  \BibitemOpen
  \bibfield  {author} {\bibinfo {author} {\bibfnamefont {L.}~\bibnamefont
  {Su}}, \bibinfo {author} {\bibfnamefont {A.~Y.}\ \bibnamefont {Piggott}},
  \bibinfo {author} {\bibfnamefont {N.~V.}\ \bibnamefont {Sapra}}, \bibinfo
  {author} {\bibfnamefont {J.}~\bibnamefont {Petykiewicz}}, \ and\ \bibinfo
  {author} {\bibfnamefont {J.}~\bibnamefont {Vuckovic}},\ }\href@noop {}
  {\bibfield  {journal} {\bibinfo  {journal} {Acs Photonics}\ }\textbf
  {\bibinfo {volume} {5}},\ \bibinfo {pages} {301} (\bibinfo {year}
  {2018})}\BibitemShut {NoStop}%
\bibitem [{\citenamefont {Birks}\ \emph {et~al.}(2015)\citenamefont {Birks},
  \citenamefont {Gris-S\'{a}nchez}, \citenamefont {Yerolatsitis}, \citenamefont
  {Leon-Saval},\ and\ \citenamefont {Thomson}}]{Birks:15}%
  \BibitemOpen
  \bibfield  {author} {\bibinfo {author} {\bibfnamefont {T.~A.}\ \bibnamefont
  {Birks}}, \bibinfo {author} {\bibfnamefont {I.}~\bibnamefont
  {Gris-S\'{a}nchez}}, \bibinfo {author} {\bibfnamefont {S.}~\bibnamefont
  {Yerolatsitis}}, \bibinfo {author} {\bibfnamefont {S.~G.}\ \bibnamefont
  {Leon-Saval}}, \ and\ \bibinfo {author} {\bibfnamefont {R.~R.}\ \bibnamefont
  {Thomson}},\ }\href {\doibase 10.1364/AOP.7.000107} {\bibfield  {journal}
  {\bibinfo  {journal} {Adv. Opt. Photon.}\ }\textbf {\bibinfo {volume} {7}},\
  \bibinfo {pages} {107} (\bibinfo {year} {2015})}\BibitemShut {NoStop}%
\bibitem [{\citenamefont {Leon-Saval}\ \emph {et~al.}(2013)\citenamefont
  {Leon-Saval}, \citenamefont {Argyros},\ and\ \citenamefont
  {Bland-Hawthorn}}]{leon20130}%
  \BibitemOpen
  \bibfield  {author} {\bibinfo {author} {\bibfnamefont {S.~G.}\ \bibnamefont
  {Leon-Saval}}, \bibinfo {author} {\bibfnamefont {A.}~\bibnamefont {Argyros}},
  \ and\ \bibinfo {author} {\bibfnamefont {J.}~\bibnamefont {Bland-Hawthorn}},\
  }\href@noop {} {\bibfield  {journal} {\bibinfo  {journal} {Nanophotonics}\
  }\textbf {\bibinfo {volume} {2}},\ \bibinfo {pages} {429} (\bibinfo {year}
  {2013})}\BibitemShut {NoStop}%
\bibitem [{\citenamefont {Resisi}\ \emph {et~al.}(2020)\citenamefont {Resisi},
  \citenamefont {Viernik}, \citenamefont {Popoff},\ and\ \citenamefont
  {Bromberg}}]{Resisi:2020}%
  \BibitemOpen
  \bibfield  {author} {\bibinfo {author} {\bibfnamefont {S.}~\bibnamefont
  {Resisi}}, \bibinfo {author} {\bibfnamefont {Y.}~\bibnamefont {Viernik}},
  \bibinfo {author} {\bibfnamefont {S.~M.}\ \bibnamefont {Popoff}}, \ and\
  \bibinfo {author} {\bibfnamefont {Y.}~\bibnamefont {Bromberg}},\ }\href
  {\doibase 10.1063/1.5136334} {\bibfield  {journal} {\bibinfo  {journal} {APL
  Photonics}\ }\textbf {\bibinfo {volume} {5}},\ \bibinfo {pages} {036103}
  (\bibinfo {year} {2020})},\ \Eprint
  {http://arxiv.org/abs/https://doi.org/10.1063/1.5136334}
  {https://doi.org/10.1063/1.5136334} \BibitemShut {NoStop}%
\bibitem [{\citenamefont {Kaina}\ \emph {et~al.}(2014)\citenamefont {Kaina},
  \citenamefont {Dupr{\'e}}, \citenamefont {Lerosey},\ and\ \citenamefont
  {Fink}}]{Kaina:2014aa}%
  \BibitemOpen
  \bibfield  {author} {\bibinfo {author} {\bibfnamefont {N.}~\bibnamefont
  {Kaina}}, \bibinfo {author} {\bibfnamefont {M.}~\bibnamefont {Dupr{\'e}}},
  \bibinfo {author} {\bibfnamefont {G.}~\bibnamefont {Lerosey}}, \ and\
  \bibinfo {author} {\bibfnamefont {M.}~\bibnamefont {Fink}},\ }\href {\doibase
  10.1038/srep06693} {\bibfield  {journal} {\bibinfo  {journal} {Scientific
  Reports}\ }\textbf {\bibinfo {volume} {4}},\ \bibinfo {pages} {6693}
  (\bibinfo {year} {2014})}\BibitemShut {NoStop}%
\bibitem [{\citenamefont {Morizur}\ \emph {et~al.}(2010)\citenamefont
  {Morizur}, \citenamefont {Nicholls}, \citenamefont {Jian}, \citenamefont
  {Armstrong}, \citenamefont {Treps}, \citenamefont {Hage}, \citenamefont
  {Hsu}, \citenamefont {Bowen}, \citenamefont {Janousek},\ and\ \citenamefont
  {Bachor}}]{Morizur:10}%
  \BibitemOpen
  \bibfield  {author} {\bibinfo {author} {\bibfnamefont {J.-F.}\ \bibnamefont
  {Morizur}}, \bibinfo {author} {\bibfnamefont {L.}~\bibnamefont {Nicholls}},
  \bibinfo {author} {\bibfnamefont {P.}~\bibnamefont {Jian}}, \bibinfo {author}
  {\bibfnamefont {S.}~\bibnamefont {Armstrong}}, \bibinfo {author}
  {\bibfnamefont {N.}~\bibnamefont {Treps}}, \bibinfo {author} {\bibfnamefont
  {B.}~\bibnamefont {Hage}}, \bibinfo {author} {\bibfnamefont {M.}~\bibnamefont
  {Hsu}}, \bibinfo {author} {\bibfnamefont {W.}~\bibnamefont {Bowen}}, \bibinfo
  {author} {\bibfnamefont {J.}~\bibnamefont {Janousek}}, \ and\ \bibinfo
  {author} {\bibfnamefont {H.-A.}\ \bibnamefont {Bachor}},\ }\href {\doibase
  10.1364/JOSAA.27.002524} {\bibfield  {journal} {\bibinfo  {journal} {J. Opt.
  Soc. Am. A}\ }\textbf {\bibinfo {volume} {27}},\ \bibinfo {pages} {2524}
  (\bibinfo {year} {2010})}\BibitemShut {NoStop}%
\bibitem [{\citenamefont {Morizur}\ \emph {et~al.}(2016)\citenamefont
  {Morizur}, \citenamefont {Labroille},\ and\ \citenamefont
  {Treps}}]{morizur2019}%
  \BibitemOpen
  \bibfield  {author} {\bibinfo {author} {\bibfnamefont {J.-F.}\ \bibnamefont
  {Morizur}}, \bibinfo {author} {\bibfnamefont {G.}~\bibnamefont {Labroille}},
  \ and\ \bibinfo {author} {\bibfnamefont {N.}~\bibnamefont {Treps}},\
  }\href@noop {} {\enquote {\bibinfo {title} {Device for processing
  light/optical radiation, method and system for designing such a device},}\ }
  (\bibinfo {year} {2016}),\ \bibinfo {note} {uS Patent 10,324,286}\BibitemShut
  {NoStop}%
\bibitem [{\citenamefont {Fontaine}\ \emph {et~al.}(2019)\citenamefont
  {Fontaine}, \citenamefont {Ryf}, \citenamefont {Chen}, \citenamefont
  {Neilson}, \citenamefont {Kim},\ and\ \citenamefont
  {Carpenter}}]{Fontaine2019}%
  \BibitemOpen
  \bibfield  {author} {\bibinfo {author} {\bibfnamefont {N.~K.}\ \bibnamefont
  {Fontaine}}, \bibinfo {author} {\bibfnamefont {R.}~\bibnamefont {Ryf}},
  \bibinfo {author} {\bibfnamefont {H.}~\bibnamefont {Chen}}, \bibinfo {author}
  {\bibfnamefont {D.~T.}\ \bibnamefont {Neilson}}, \bibinfo {author}
  {\bibfnamefont {K.}~\bibnamefont {Kim}}, \ and\ \bibinfo {author}
  {\bibfnamefont {J.}~\bibnamefont {Carpenter}},\ }\href {\doibase
  10.1038/s41467-019-09840-4} {\bibfield  {journal} {\bibinfo  {journal}
  {Nature Communications}\ }\textbf {\bibinfo {volume} {10}},\ \bibinfo {pages}
  {1865} (\bibinfo {year} {2019})}\BibitemShut {NoStop}%
\bibitem [{\citenamefont {Labroille}\ \emph {et~al.}(2014)\citenamefont
  {Labroille}, \citenamefont {Denolle}, \citenamefont {Jian}, \citenamefont
  {Genevaux}, \citenamefont {Treps},\ and\ \citenamefont
  {Morizur}}]{Labroille:14}%
  \BibitemOpen
  \bibfield  {author} {\bibinfo {author} {\bibfnamefont {G.}~\bibnamefont
  {Labroille}}, \bibinfo {author} {\bibfnamefont {B.}~\bibnamefont {Denolle}},
  \bibinfo {author} {\bibfnamefont {P.}~\bibnamefont {Jian}}, \bibinfo {author}
  {\bibfnamefont {P.}~\bibnamefont {Genevaux}}, \bibinfo {author}
  {\bibfnamefont {N.}~\bibnamefont {Treps}}, \ and\ \bibinfo {author}
  {\bibfnamefont {J.-F.}\ \bibnamefont {Morizur}},\ }\href {\doibase
  10.1364/OE.22.015599} {\bibfield  {journal} {\bibinfo  {journal} {Opt.
  Express}\ }\textbf {\bibinfo {volume} {22}},\ \bibinfo {pages} {15599}
  (\bibinfo {year} {2014})}\BibitemShut {NoStop}%
\bibitem [{\citenamefont {Brandt}\ \emph {et~al.}(2020)\citenamefont {Brandt},
  \citenamefont {Hiekkam\"{a}ki}, \citenamefont {Bouchard}, \citenamefont
  {Huber},\ and\ \citenamefont {Fickler}}]{Brandt:20}%
  \BibitemOpen
  \bibfield  {author} {\bibinfo {author} {\bibfnamefont {F.}~\bibnamefont
  {Brandt}}, \bibinfo {author} {\bibfnamefont {M.}~\bibnamefont
  {Hiekkam\"{a}ki}}, \bibinfo {author} {\bibfnamefont {F.}~\bibnamefont
  {Bouchard}}, \bibinfo {author} {\bibfnamefont {M.}~\bibnamefont {Huber}}, \
  and\ \bibinfo {author} {\bibfnamefont {R.}~\bibnamefont {Fickler}},\ }\href
  {\doibase 10.1364/OPTICA.375875} {\bibfield  {journal} {\bibinfo  {journal}
  {Optica}\ }\textbf {\bibinfo {volume} {7}},\ \bibinfo {pages} {98} (\bibinfo
  {year} {2020})}\BibitemShut {NoStop}%
\bibitem [{\citenamefont {L{\'o}pez-Pastor}\ \emph {et~al.}(2019)\citenamefont
  {L{\'o}pez-Pastor}, \citenamefont {Lundeen},\ and\ \citenamefont
  {Marquardt}}]{lopezpastor2019}%
  \BibitemOpen
  \bibfield  {author} {\bibinfo {author} {\bibfnamefont {V.~J.}\ \bibnamefont
  {L{\'o}pez-Pastor}}, \bibinfo {author} {\bibfnamefont {J.~S.}\ \bibnamefont
  {Lundeen}}, \ and\ \bibinfo {author} {\bibfnamefont {F.}~\bibnamefont
  {Marquardt}},\ }\href@noop {} {\enquote {\bibinfo {title} {Arbitrary optical
  wave evolution with fourier transforms and phase masks},}\ } (\bibinfo {year}
  {2019}),\ \Eprint {http://arxiv.org/abs/1912.04721} {arXiv:1912.04721
  [quant-ph]} \BibitemShut {NoStop}%
\bibitem [{\citenamefont {Beenakker}(1997)}]{Bennaker_1997}%
  \BibitemOpen
  \bibfield  {author} {\bibinfo {author} {\bibfnamefont {C.~W.~J.}\
  \bibnamefont {Beenakker}},\ }\href {\doibase 10.1103/RevModPhys.69.731}
  {\bibfield  {journal} {\bibinfo  {journal} {Rev. Mod. Phys.}\ }\textbf
  {\bibinfo {volume} {69}},\ \bibinfo {pages} {731} (\bibinfo {year}
  {1997})}\BibitemShut {NoStop}%
\bibitem [{\citenamefont {Tulino}\ \emph {et~al.}(2004)\citenamefont {Tulino},
  \citenamefont {Verd{\'u}} \emph {et~al.}}]{tulino2004}%
  \BibitemOpen
  \bibfield  {author} {\bibinfo {author} {\bibfnamefont {A.~M.}\ \bibnamefont
  {Tulino}}, \bibinfo {author} {\bibfnamefont {S.}~\bibnamefont {Verd{\'u}}},
  \emph {et~al.},\ }\href@noop {} {\bibfield  {journal} {\bibinfo  {journal}
  {Foundations and Trends{\textregistered} in Communications and Information
  Theory}\ }\textbf {\bibinfo {volume} {1}},\ \bibinfo {pages} {1} (\bibinfo
  {year} {2004})}\BibitemShut {NoStop}%
\bibitem [{\citenamefont {Goetschy}\ and\ \citenamefont
  {Stone}(2013)}]{Goetschy13}%
  \BibitemOpen
  \bibfield  {author} {\bibinfo {author} {\bibfnamefont {A.}~\bibnamefont
  {Goetschy}}\ and\ \bibinfo {author} {\bibfnamefont {A.~D.}\ \bibnamefont
  {Stone}},\ }\href {\doibase 10.1103/PhysRevLett.111.063901} {\bibfield
  {journal} {\bibinfo  {journal} {Phys. Rev. Lett.}\ }\textbf {\bibinfo
  {volume} {111}},\ \bibinfo {pages} {063901} (\bibinfo {year}
  {2013})}\BibitemShut {NoStop}%
\end{thebibliography}%

\end{document}